\documentclass[osa,twocolumn,showpacs,floatfix]{revtex4}

\usepackage{graphicx}%
\usepackage{here}

\newcommand\pictc[5]{\begin{figure}
                     \centerline{
                     \includegraphics[width=#1\columnwidth]{#3}}
           \vspace{-0.2cm}
                   \protect\caption{\protect\label{fig:#4} #5
                   \vspace{-0.4cm}}
                    \end{figure}            }
\newcommand\pict[4][1.]{\pictc{#1}{!tb}{#2}{#3}{#4}}

\newcounter{Fig}

\begin{document}
\begin{sloppy}

\title{Soliton stripes in two-dimensional nonlinear photonic lattices}

\author{Dragomir Neshev}
\author{Yuri S. Kivshar}
\affiliation{Nonlinear Physics Group, Research School of Physical Sciences and Engineering, Australian National University, Canberra ACT 0200, Australia}
%\homepage{http://www.rsphysse.anu.edu.au/nonlinear}

\author{Hector Martin}
\author{Zhigang Chen}
\affiliation{Department of Physics and Astronomy, San Francisco State University, San Francisco, California 94132 and Teda College, Nankai University, Tianjing, China}

\begin{abstract}
We study experimentally the interaction of a soliton with a nonlinear lattice. We observe the formation of a novel type of composite soliton created by strong coupling of mutually incoherent periodic and localized beam components. By imposing an initial transverse momentum on the soliton stripe, we observe the effect of lattice compression and deformation.
\end{abstract}

% (190.0190) Nonlinear optics
% (190.4420) Nonlinear optics, transverse effects in

\maketitle

The study of nonlinear light propagation in periodic photonic structures recently attracted a strong interest due to the unique possibility to observe an interplay between the effects of nonlinearity and periodicity. In particular, a periodic modulation of the refractive index modifies the linear spectrum and wave diffraction and, consequently, strongly affects the nonlinear propagation and localization of light~\cite{book}. Recently, many nonlinear effects, including the formation of lattice solitons, were demonstrated experimentally for one-~\cite{Fleischer:2003-23902:PRL,Neshev:2003-710:OL} and two-~\cite{Fleischer:2003-147:Nature,Martin:arXive} dimensional optically-induced photonic lattices. The main idea of these studies, first suggested theoretically by Efremidis {\em et al.}~\cite{efrem}, is to use a photorefractive crystal with a strong electro-optic anisotropy to create an optically-induced lattice with a polarization orthogonal to that of a probe beam. Due to the substantial difference in electro-optic coefficients for the two polarizations, the material nonlinearity experienced by the lattice is at least an order of magnitude weaker than that experienced by the probe beam. Thus the lattice-forming waves propagate {\em in the linear regime}.

On the other hand, {\em nonlinear lattices} formed by solitons, were also recently demonstrated experimentally in parametric processes~\cite{Minardi:2000-326:OL} and in photorefractive crystals with both coherent~\cite{Petter:2003-438:OL,Petrovic:arXive} and partially incoherent light~\cite{Klinger:2001-271:OL,Chen:2002-2019:OL}. Such nonlinear lattices offer unique possibilities for the study of nonlinear effects in periodic systems, allowing to expand the concept of optically-induced lattices beyond the limits of weak material nonlinearity~\cite{Desyatnikov:nlin.PS/0304008}. In particular, it was shown theoretically that strong interaction of a periodic lattice with a probe beam, through the nonlinear cross-phase modulation effect, facilitates the formation of a novel type of composite optical soliton, where one of the components creates a nonlinear periodic structure that traps and localizes the other~\cite{Desyatnikov:nlin.PS/0304008}. Observation of nonlinear light localization of this type does not require strong anisotropy of nonlinear properties of the medium, as both the periodic pattern and the localized beam, although mutually incoherent, are of the same polarization. In a bulk medium, one-dimensional solitons, localized or periodic, are represented by quasi-one-dimensional entities, infinitely extended along the second transverse dimension. Such objects, however, suffer from transverse modulational instability~\cite{book}. This transverse instability can be avoided if the second, homogeneous dimension is periodically modulated with a period shorter then the minimal instability length~\cite{Barthelemy}. Therefore, in order to investigate the nonlinear interaction between periodic and localized beam one need to consider a hybrid interaction of a stripe and a two-dimensional nonlinear lattice. 

%--------------------------------------------------------------------------
\pict[0.92]{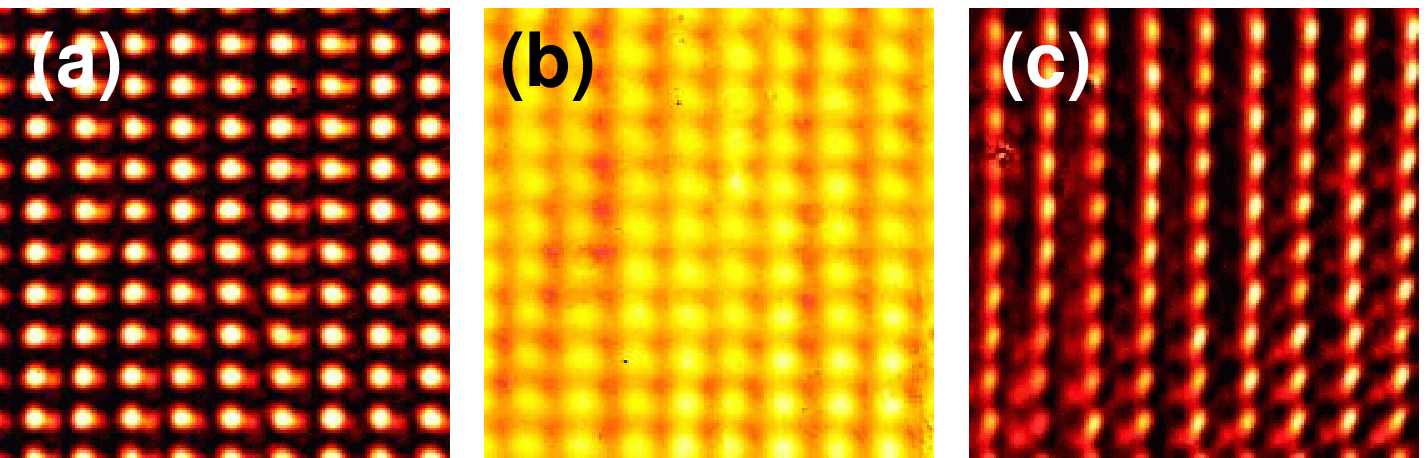}{excitation_lattice}{Creation of a two-dimensional optically-induced nonlinear lattice: (a) input intensity distribution; (b) linear diffraction at the output face of the crystal; (c) formation of a soliton lattice at the crystal output.}
%--------------------------------------------------------------------------

Thus, the purpose of this Letter is twofold: (i) to present the first experimental observation of novel types of nonlinear coupled soliton states in optically-induced nonlinear lattices, and (ii) to study experimentally different regimes of nonlinear interaction between a soliton stripe and a two-dimensional soliton lattice. Interaction of the stripe with the soliton lattice can lead to a break-up of the stripe due to the hybrid interaction forces~\cite{DelRe:2000-560:OL}, and the formation of a fully periodic state (a soliton array) in the direction of the stripe. Respectively, this soliton array can be coupled to the two-dimensional soliton lattice. We explore the stripe-lattice interaction when the stripe is tilted with respect to the lattice orientation, thus breaking the interaction symmetry. Additionally, we investigate the interaction when the stripe has an initial transverse momentum relative to the lattice and demonstrate lattice compression and deformation due to the interaction with the stripe.

%--------------------------------------------------------------------------

First of all, we create a stable and regular square lattice by illuminating an amplitude mask with partially incoherent light and imaging it onto the input face of a biased photorefractive crystal (SBN:60, 8mm long). Details of the experimental setup have been discussed earlier in Ref.~\cite{Chen:2002-2019:OL}. The input intensity distribution corresponding to the generated lattice is shown in Fig.~\ref{fig:excitation_lattice}(a). The degree of spatial coherence is about 40$\mu$m, such that the lattice pattern becomes fairly uniform after linear propagation through the crystal [see Fig.~\ref{fig:excitation_lattice}(b)] but a regular soliton lattice (with the lattice spacing of 36.7$\mu$m) is observed for a biasing field of 2900~V/cm [see Fig.~\ref{fig:excitation_lattice}(c)].

To study the stripe interaction with the two-dimensional nonlinear lattice, we launch another, coherent beam (but mutually incoherent with the lattice) and focus it by a cylindrical lens onto the input face of the crystal. The resulting stripe has a full width half maximum (FWHM) $\sim 20\mu$m along the $x$-axis, has homogeneous intensity along $y$-axis, and it propagates parallel to the lattice along the $z$-direction. First, we orient the stripe to be perpendicular to the c-axis of the crystal [Fig.~\ref{fig:parallel_stripe}(a)] and to coincide with a single column of lattice sites. The input intensities of both beams are set in such a way that the generation of a stripe soliton and a soliton lattice can be achieved for a fixed voltage applied to the crystal. In our case, the initial lattice has an intensity 1.2 times higher than that of the stripe. Without the applied electric field, the stripe diffracts homogeneously to a FWHM of 100$\mu$m [Fig.~\ref{fig:parallel_stripe}(b)]. When voltage is applied to the crystal (electric field 2900~V/cm), we observe self-trapping effect for the stripe alone and the formation of a quasi-one-dimensional spatial soliton [Fig.~\ref{fig:parallel_stripe}(c)]. Due to the saturation of the photorefractive nonlinearity development of transverse break-up of the stripe is not observed at the end of the crystal. Because of diffusion present in a photorefractive medium, the output position of the stripe is observed to be shifted by $\sim 43\mu$m to the right, along the $x$-axis. However, when the stripe propagates simultaneously with the nonlinear lattice, formation of a {\it coupled state} is observed, in which the stripe trapped by the lattice is not affected by the diffusion [Fig.~\ref{fig:parallel_stripe}(d,e)], and preserves its initial position. We note that the diffusion also bends the lattice, but to a much smaller extent, since the lattice covers a significant part of the crystal.

%--------------------------------------------------------------------------
%--------------------------------------------------------------------------
\pict[0.96]{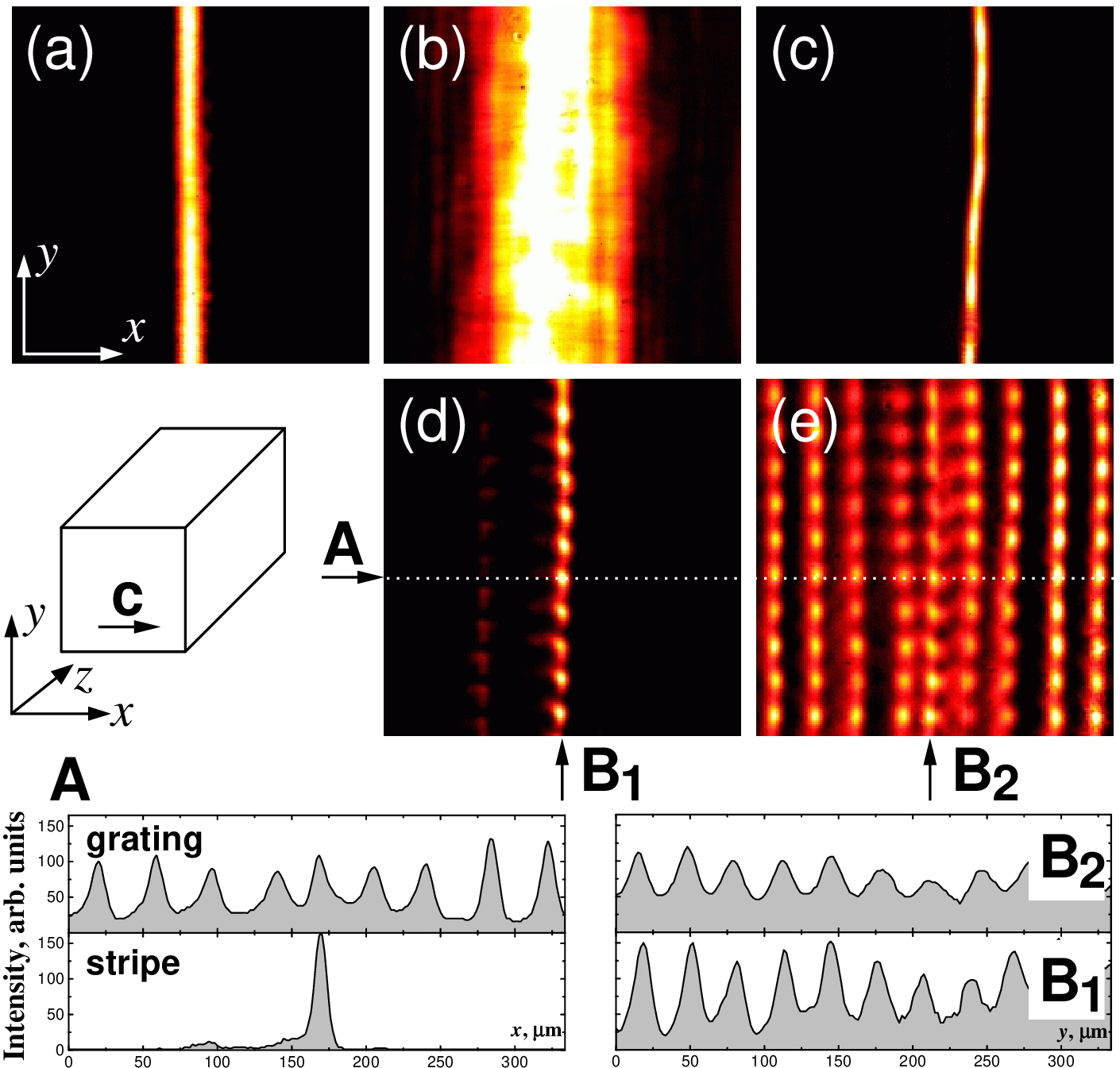}{parallel_stripe}{Upper row: propagation of a stripe oriented perpendicular to the crystal c-axis. (a) Input; (b) linear diffraction at the output; (c) output after nonlinear propagation (applied field 2900~V/cm). Lower rows: formation of a composite lattice-stripe soliton. (d,e) Output intensity distribution of the stripe and lattice components, respectively,  after their simultaneous propagation in the crystal. Transverse intensity distributions of the coupled state are shown below for the horizontal ({\sf A}) and vertical (${\sf B_1}$ and ${\sf B_2}$) cross-sections.}
%--------------------------------------------------------------------------
%--------------------------------------------------------------------------
\pict[0.96]{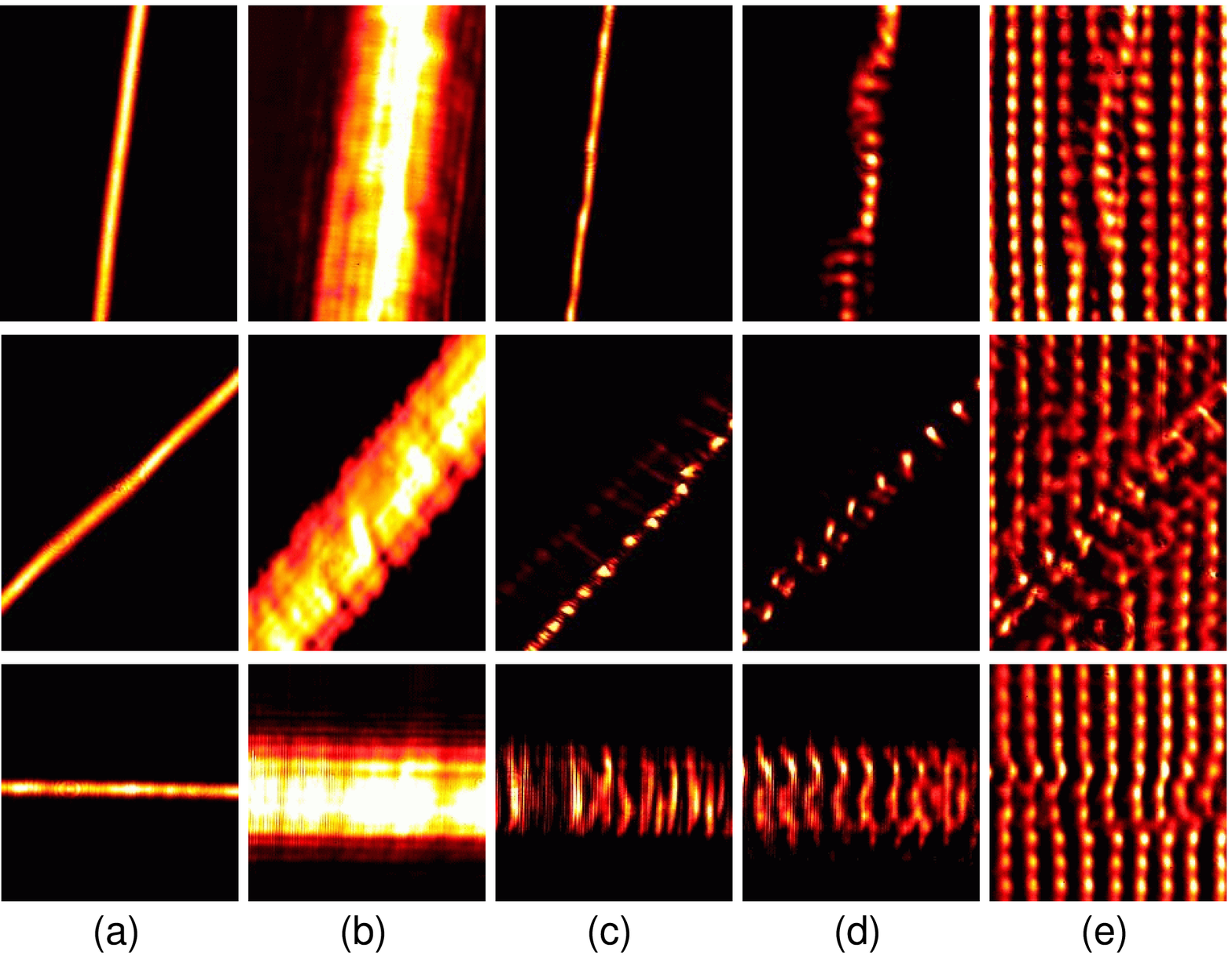}{tilted_stripe}{ Interaction of a lattice with a tilted stripe. Top row: small tilting ($\sim 7^\circ$) from the vertical $y$ direction; a soliton is formed. Middle row: $45^\circ$ angle from the $y$-axis. Bottom row: $90^\circ$ angle from the $y$-axis. (a-e) Shown is the same as in
Fig.~\ref{fig:parallel_stripe}(a-e). Applied field 2900~V/cm.}
%--------------------------------------------------------------------------

The formation of a composite state that consists of a localized and periodic components strongly coupled by the cross-phase modulation can be clearly seen from the analysis of the intensity profiles of the vertical and horizontal cross-sections of the two components, as shown in the plots ({\sf A}) and (${\sf B_1}$ and ${\sf B_2}$) of Fig.~\ref{fig:parallel_stripe}. In the horizontal direction ({\sf A}), the lattice is deformed by the presence of the second beam. As predicted theoretically by Desyatnikov {\em et al.}~\cite{Desyatnikov:nlin.PS/0304008}, the nonlinear interaction between the soliton array and a localized beam can lead to a deformation of the array and the formation of a novel composite band-gap soliton. 
%where one of the components creates a periodic nonlinear lattice which traps the other component in the form of a lattice or gap soliton. 
In the experiment, we observe that, when both components propagate simultaneously, the central and two neighboring lattice sites come closer together and decrease their intensities, thus indicating the formation of a coupled composite state. At the same time, the stripe becomes periodically modulated by the lattice along the vertical direction ({\sf B}), so that each intensity maxima is trapped on a single lattice site. As discussed earlier, such induced modulation suppresses the transverse instability of the stripe, observed in the case of quasi-one-dimensional periodic structures~\cite{Neshev:nlin.PS/0307053}.

%--------------------------------------------------------------------------

When the stripe is located initially between two neighboring lattice sites, the resulting coupled state is found to be highly unstable and very sensitive to the initial alignment. Either a strong deformation of the lattice or a transition to the coupled state presented above is observed. To verify the position sensitivity for the formation of a bound state, we tilted the stripe at a small angle to the vertical $y$-axis, distorting the initial symmetry [Fig.~\ref{fig:tilted_stripe}(top row)]. Tilting the stripe away from its vertical position is expected to affect its nonlinear propagation due to the inherent nonlinear anisotropy of the photorefractive crystal. For a small tilt of $\sim7^\circ$, however, the stripe still forms a one-dimensional soliton preserving the tilting angle at the output. The simultaneous propagation of the stripe and the lattice, however, exhibits a strong interaction, which tends to restore the vertical symmetry [Fig.~\ref{fig:tilted_stripe}(top row)(d,e)]. Therefore, along the vertical direction, a ``ladder'' of different coupled states is observed, where the position of the maximum of the stripe experiences {\it discrete jumps} from one lattice site to the next one. In the intermediate regions, the lattice is strongly distorted.

%--------------------------------------------------------------------------
\pict[0.6]{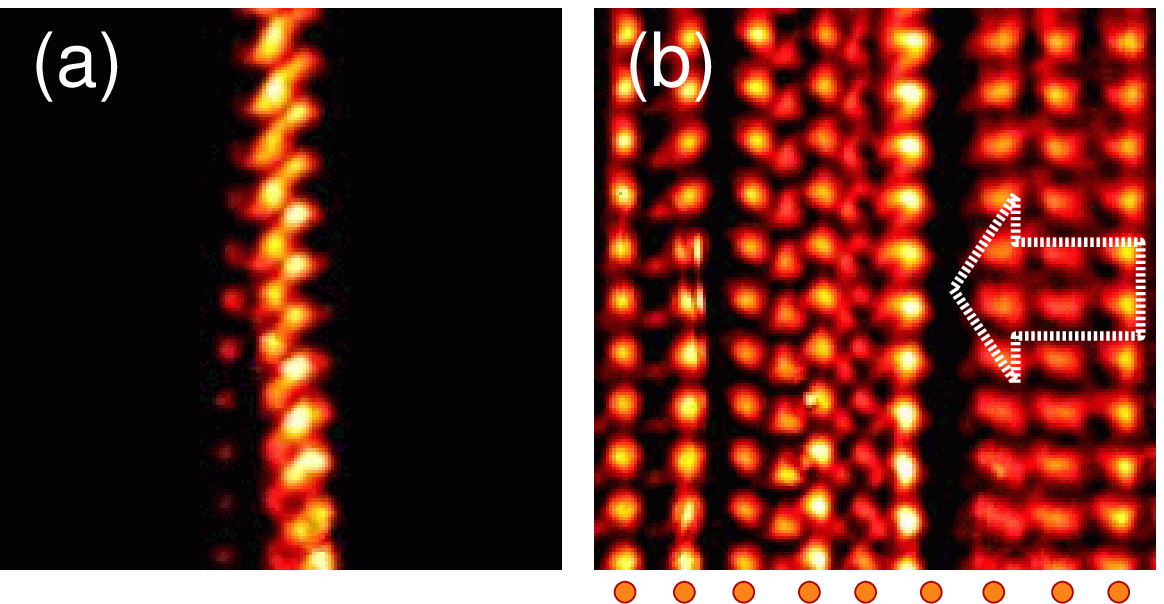}{compression_wave}{Compression and deformation of the lattice by a vertical stripe propagating under a small angle. (a) Stripe at the output after interaction with a lattice. (b) Lattice at the output. The dots under the image show the positions of the unperturbed lattice columns, and the arrow indicates the direction of the initial transverse momentum of the stripe.}
%--------------------------------------------------------------------------

Due to the anisotropic photorefractive nonlinearity, a larger initial tilt of the stripe seems to afflict the formation of a stripe soliton. Indeed, as shown in Fig.~\ref{fig:tilted_stripe}(middle row), the $45^\circ$ tilted stripe does not fully self-focus and, instead, it breaks-up even when propagating alone. Interaction of such a beam with the lattice leads to its strong deformation, as shown in Fig.~\ref{fig:tilted_stripe}(middle row)(e). An interesting example of the effect of anisotropic nonlinearity is observed when the initial stripe is oriented horizontally, along the c-axis of the crystal [see Fig.~\ref{fig:tilted_stripe}(bottom row)]. For this case, the horizontal break-up of the diffracted stripe leads to the formation of vertical filaments [Fig.~\ref{fig:tilted_stripe}(bottom row)(c)]. These filaments may be considered as low-intensity vertical stripes which can be trapped by the lattice and aligned along its columns [Fig.~\ref{fig:tilted_stripe}(bottom row)(d)]. The lattice itself is weakly distorted and generally preserves its symmetry [Fig.~\ref{fig:tilted_stripe}(bottom row)(e)].

To study the nonlinear interaction between the lattice and a soliton stripe moving in the $x-z$ plane, we inclined the propagation direction of the stripe by $\sim 0.3^\circ$ with respect to that of the lattice (i.e., the $z$-direction), thus imposing an initial transverse momentum of the stripe relative to the lattice. In the linear regime, this should lead to the  discreteness-induced effects in the stripe propagation. However, in the nonlinear regime, such an initial inclination leads to dragging of the entire lattice column which causes overall lattice compression in the direction of the transverse momentum. As seen in Fig.~\ref{fig:compression_wave}(a), due to the interaction, the stripe beam splits into two columns of periodic peaks that are half-period shifted in the vertical direction. On the other hand the lattice is compressed toward the direction of the stripe momentum. This lattice deformation creates an additional lattice column, and the first compressed lattice columns are shifted vertically approximately to a half of the lattice period
%, as shown in 
[Fig.~\ref{fig:compression_wave}(b)].

In conclusion, we have observed, for the first time to our knowledge, the formation of a novel type of composite soliton in the form of a nonlinear state of periodic and localized field components mutually incoherent but strongly coupled via the cross-phase modulation. Such a coupled state is a robust object, and it is preserved under distortions of its symmetry. Imposing an initial transverse momentum to the stripe leads to the deformation and the compression of the two-dimensional optically-induced photonic lattice.

D.N. thanks the Department of Physics and Astronomy of the San Francisco State University for its hospitality. The work was partially supported by the Australian Research Council, the U.S. Air Force Office of Scientific Research, and Research Corporation.

\vspace{-0.5cm}
%\newpage
%--------------------------------------------------------------------------

%\pagebreak

%\newpage
%\newpage
%\newpage
%\newpage
%\pagebreak
%\pagebreak
%\pictFig{fig1}
%\pictFig{fig2}
%\pictFig{fig3}
%\pictFig{fig4}

%==========================================================================
\end{sloppy}
\end{document}